\documentclass[aps,prd,showpacs,floatfix,preprint]{revtex4}
\usepackage{amsmath,bm}
\usepackage{graphicx}
\usepackage{epsfig}
\begin{document}

\title{Shear viscosity and the nucleation of antikaon condensed matter in 
protoneutron stars}

\author{Sarmistha Banik, Rana Nandi and Debades Bandyopadhyay}
\affiliation{Astroparticle Physics and Cosmology Division,
Saha Institute of Nuclear Physics, 1/AF Bidhannagar, 
Kolkata-700064, India}

\begin{abstract}
We study shear viscosities of different species in hot and neutrino-trapped
dense matter relevant to protoneutron stars. It is found that the shear
viscosities of neutrons, protons and electrons in neutrino-trapped matter are
of the same orders of magnitude as the corresponding shear viscosities
in neutrino-free matter. Above all, the shear viscosity due to neutrinos is 
higher by several orders of magnitude than that of other species in 
neutrino-trapped matter. 
 
Next we investigate the effect of shear viscosity in particular, neutrino shear
viscosity on the thermal nucleation rate of droplets of antikaon condensed 
matter in protoneutron stars. The first-order phase transition from hadronic 
matter to antikaon condensed matter is driven by the thermal nucleation 
process. 
We compute the equation of state used for the calculation of shear viscosity 
and thermal nucleation time within the relativistic mean field model. Neutrino
shear viscosity enhances the prefactor in the nucleation rate by 
several orders of magnitude compared with the $T^4$ approximation of earlier 
calculations. 
Consequently the thermal nucleation time in the $T^4$ approximation 
overestimates our result. Furthermore, the thermal nucleation of
an antikaon droplet might be possible in neutrino-trapped matter before neutrino diffusion
takes place. 
\end{abstract}

\maketitle

\section{Introduction}
A first order phase transition from nuclear matter to some exotic form of 
matter might be possible in (proto)neutron stars. It could be either
a nuclear to quark matter transition or a first order pion/kaon condensation.
Consequently, it might have tremendous implications for compact stars 
\cite{Gle} and supernova explosions \cite{Sag}. Here the focus is the first
order phase transition proceeding through the thermal nucleation of a new phase
in particular, antikaon condensed phase in hot and neutrino-trapped matter. 
After the pioneering work by Kaplan and Nelson on   
antikaon ($K^-$ meson) condensation in dense baryonic matter formed in heavy 
ion collisions as well as in neutron stars 
\cite{Kap}, several groups pursued the problem of antikaon 
condensation in (proto)neutron stars 
\cite{Bro,Tho,Ell,Lee,Pra97,Gle99,Kno,Sch,Pal,Bani1,Bani2,Bani3,Bani4,Pons,Bani5}. In most cases, the
phase transition was studied using either Maxwell construction or 
Gibbs' rules for phase equilibrium coupled with global baryon number and 
charge conservation \cite{Gle92}. The first order phase transition driven by
nucleation of antikaon condensed phase was considered in a few cases 
\cite{Nor,Bani7}. In particular, the calculation of Ref.\cite{Bani7} dealt with
the role of shear viscosity on the the thermal nucleation of antikaon condensed
phase in hot and neutrino-free compact stars \cite{Bani7}. It is to be noted 
here that the first order phase transition through the thermal nucleation of 
quark matter droplets was also investigated in (proto)neutron stars 
\cite{Nor,San,Sat,Hei,Bom,Min,Bom2} using the homogeneous 
nucleation theory of Langer \cite{Nor,San,Lan}. The thermal nucleation is an 
efficient process than the quantum nucleation at high temperatures 
\cite{Bom,Bom2}.

We adopt the homogeneous nucleation theory of Langer \cite{Lan,Tur} for the 
thermal nucleation of antikaon condensed phase. Nuclear matter would be 
metastable near the phase transition point due to sudden change in state 
variables. In this case thermal and quantum fluctuations are important. 
Droplets of antikaon condensed matter are formed 
because of thermal fluctuations in the metastable nuclear matter. 
Droplets of the new and stable phase which are bigger than a critical radius,
will survive and grow. The transportation of latent heat from the surface of 
the droplet into the metastable phase favours a critical size droplet to grow 
further. This heat transportation could be achieved through the thermal 
dissipation and viscous damping \cite{Tur,Las,Raj}. 

A parametrised form of the shear viscosity was used in 
earlier calculations of the nucleation of quark matter \cite{Bom}. 
Recently, the influence of thermal conductivity and
shear viscosity on the thermal nucleation time was studied in a first-order
phase transition from nuclear to antikaon condensed matter in hot neutron
stars \cite{Bani7}. The shear viscosity due to neutrinos was not 
considered in that calculation. It would be worth 
studying the effect of shear viscosity on the thermal 
nucleation rate of droplets of antikaon condensed matter in neutrino-trapped
matter relevant for protoneutron stars. Besides shear viscosities due to
neutrons, protons and electrons, this involves the contribution of neutrinos to
the total shear viscosity. Shear viscosities of pure neutron and neutron star 
matter were calculated by several groups \cite{Flo1,Flo2,Cut,Ben,Yak,Glam}. 
We also performed the calculation of shear viscosity in neutron star matter 
using the equation of state (EoS) derived from relativistic field theoretical 
models \cite{Bani6}. Transport properties of degenerate neutrinos in dense 
matter were estimated by Goodwin and Pethick \cite{Good}. 

We organise the paper in the following way. We describe models for 
EoS, shear viscosities of different species including neutrinos and the
calculation of thermal nucleation rate in Sec. II.
Results of this calculation are discussed in Sec. III. Sec. IV gives the
summary and conclusions.

\section{Formalism}
The knowledge of the EoS for nuclear as well as antikaon condensed phases is 
essential for the computation of shear viscosity and thermal nucleation rate. 
We consider a first-order phase transition from the charge neutral and 
$\beta$-equilibrated nuclear matter to $K^-$ condensed matter in a protoneutron 
star. Those two phases are composed of neutrons, protons, electrons, electron 
type neutrinos and of $K^-$ mesons only in the antikaon condensed phase. 
Both phases are governed by baryon number conservation and charge 
neutrality conditions \cite{Gle92}. Relativistic field theoretical models are 
used to describe the EoS in nuclear and antikaon condensed phases.  
The baryon-baryon interaction is mediated by the exchange of $\sigma$, $\omega$
and $\rho$ and given by the Lagrangian density 
\cite{Sch,walecka,serot, glendenning,Bani1}    
\begin{eqnarray}
{\cal L}_N &=& \sum_{B=n,p} \bar\psi_{B}\left(i\gamma_\mu 
{\partial}^{\mu} - m_B
+ g_{\sigma B} \sigma - g_{\omega B} \gamma_\mu \omega^\mu
- g_{\rho B}
\gamma_\mu{\mbox{\boldmath t}}_B \cdot
{\mbox{\boldmath $\rho$}}^\mu \right)\psi_B\nonumber\\
&& + \frac{1}{2}\left( \partial_\mu \sigma\partial^\mu \sigma
- m_\sigma^2 \sigma^2\right) - U(\sigma) \nonumber\\
&& -\frac{1}{4} \omega_{\mu\nu}\omega^{\mu\nu}
+\frac{1}{2}m_\omega^2 \omega_\mu \omega^\mu
- \frac{1}{4}{\mbox {\boldmath $\rho$}}_{\mu\nu} \cdot
{\mbox {\boldmath $\rho$}}^{\mu\nu}
+ \frac{1}{2}m_\rho^2 {\mbox {\boldmath $\rho$}}_\mu \cdot
{\mbox {\boldmath $\rho$}}^\mu ~.
\end{eqnarray}
The scalar self-interaction term \cite{Sch,glendenning,boguta} is 
\begin{equation}
U(\sigma)~=~\frac13~g_1~m_N~(g_{\sigma N}\sigma)^3~+~ \frac14~g_2~
(g_{\sigma N}\sigma)^4~,
\end{equation}
The effective nucleon mass is given by $m_B^* = m_B - g_{\sigma B} \sigma$, 
where $m_B$ is the vacuum baryon mass.
The Lagrangian density for (anti)kaons in the minimal coupling scheme is given 
by
\cite{Pal,Bani1,Bani2,Gle99}
\begin{equation}
{\cal L}_K = D^*_\mu{\bar K} D^\mu K - m_K^{* 2} {\bar K} K ~,
\end{equation}
where the covariant derivative is
$D_\mu = \partial_\mu + ig_{\omega K}{\omega_\mu}
+ i g_{\rho K}
{\mbox{\boldmath t}}_K \cdot {\mbox{\boldmath $\rho$}}_\mu$ and
the effective mass of (anti)kaons is
$m_K^* = m_K - g_{\sigma K} \sigma$.
For s-wave (${\bf p}=0$) condensation, the in-medium energy of $K^{-}$ mesons 
is given by
\begin{equation}
\omega_{K^{-}} =  m_K^{*} - \left( g_{\omega K} \omega_0
+ \frac {1}{2} g_{\rho K} \rho_{03} \right)~.
\end{equation}
The condensation in 
neutrino-trapped matter sets in when the chemical potential of $K^-$ mesons  
($\mu_{K^-} = \omega_{K^-}$) is equal to the difference between the electron 
chemical 
potential ($\mu_e$) and neutrino chemical potential ($\mu_{\nu_e}$) i.e. 
$\mu_{K^-} = \mu_e - \mu_{\nu_e}$. 
The critical droplet of antikaon condensed matter is in 
total phase equilibrium with the metastable nuclear matter. The mixed phase is 
governed by Gibbs' phase rules along with global baryon number conservation
and charge neutrality \cite{Gle92}. 

We adopt the bulk approximation \cite{San} which does not consider the 
variation of the meson fields with position inside the droplet. 
We solve equations of motion self-consistently in the
mean field approximation \cite{walecka} and find effective masses and Fermi 
momenta of baryons. Pressures in neutrino-trapped nuclear matter ($P^N$) 
and antikaon condensed matter ($P^K$) are given by Ref.\cite{Bani1}. Here
we calculate zero temperature equations of state because it was noted earlier
that the temperature of a few tens of MeV did not modify the EoS considerably 
\cite{Bani7}.

Next we discuss the calculation of shear viscosity in protoneutron star matter.
It was noted in earlier calculations that the main contributions to the total 
shear viscosity in neutron star matter came from electrons, the lightest 
charged particles, and neutrons, the most abundant particles 
\cite{Flo1,Flo2,Cut,Yak,Glam}. Neutrinos are trapped in protoneutron stars and
their contribution might be significant in transport coefficients such as
shear viscosity. In principle, we may calculate shear viscosities for different
particle species ($n$, $p$, $e$ and $\nu_e$) in 
neutrino-trapped matter using coupled Boltzmann transport equations 
\cite{Flo2,Yak,Bani6}. We can immediately write a set of relations between 
effective relaxation times ($\tau$) and collision frequencies ($\nu_{ij}^{(')}$)
following the Ref.\cite{Bani6}
\begin{equation}
\sum_{i,j=n,p,e,\nu_e} (\nu_{ij}\tau_i + \nu'_{ij}\tau_j) = 1~,
\end{equation}
which can be cast into a $4\times4$ matrix. However, solving the matrix 
equation becomes
a problem because the relaxation time for neutrinos is much larger than those of
other species as it is evident from Fig. 3. So we calculate shear viscosities 
due to neutrons, protons and 
electrons in neutrino-trapped matter from the above matrix equation as it was 
done in Ref.\cite{Bani6}. In this case,  
the effects of the exchange of transverse plasmons in the collisions of charged
particles are included \cite{Yak,Yak2}. The knowledge of 
nucleon-nucleon scattering cross sections obtained from the Dirac-Brueckner 
approach \cite{Mach1,Mach2} is exploited for the calculation of neutron and 
proton shear viscosities \cite{Yak,Bani6,Bai}. On the other hand, we estimate
the neutrino shear viscosity using the prescription of Goodwin and Pethick
\cite{Good},
\begin{equation}
\eta_{\nu} = \frac{1}{5} n_{\nu} p_{\nu} c \tau_{\nu} \left[\frac{\pi^2}{12} 
+ \lambda_{\eta}\sum_{k=odd} \frac{2(2k+1)}{k^2(k+1)^2[k(k+1)-2\lambda_{\eta}]}\right]~. 
\end{equation}
For neutrino shear viscosity, we only consider scattering processes involving
neutrinos and other species. Various quantities in Eq.(6) are explained in
Ref.\cite{Good} and given below.
The neutrino relaxation time ($\tau_{\nu}$) is,
\begin{eqnarray}
 \tau_{\nu}^{-1}&=&\sum_{i=n,p,e}\tau_{{\nu}i}^{-1},\hspace{0.5cm}~,\\
 \tau_{{\nu}i}^{-1}&=&\frac{E_{F_i}^2(k_BT)^2}{64\pi^2}<I^i>~,
\end{eqnarray}
and $\lambda_\eta$ is defined as
\begin{eqnarray}
 \lambda_\eta &=&\tau_{\nu}\sum_i\lambda_\eta^i\tau_{{\nu}i}^{-1}\\
 \lambda_\eta^i&=&\frac{ \int d\Omega_2d\Omega_3d\Omega_4\,\frac{1}{2}\left[3(\hat{{\bf p}_1}\cdot\hat{{\bf p}_3})^2-1\right]\left<|M|^2\right>_i
                   \delta({\bf p}_1+{\bf p}_2-{\bf p}_3-{\bf p}_4)}{\int d\Omega_2d\Omega_3d\Omega_4\left<|M|^2\right>_i\delta({\bf p}_1+{\bf p}_2-{\bf p}_3-{\bf p}_4)}\\
               &=&1-\frac{3}{2p_{\nu}^2}\frac{I_1^i}{<I^i>}+\frac{3}{8p_\nu^4}\frac{I_2^i}{<I^i>}~,\label{etai}
\end{eqnarray}{\small
\begin{eqnarray}
 <I^i>&=&\frac{8G_F^2\pi p_\nu}{3}\Bigg[4C_{V_i}C_{A_i}\left(\frac{p_\nu}{E_i}\right)+(C_{V_i}^2+C_{A_i}^2)
    \left\{ 3 + \left(\frac{p_i}{E_i}\right)^2+\frac{2}{5}\left(\frac{p_\nu}{E_i}\right)^2\right\}\nonumber\\
    &&-\left(\frac{m_i}{E_i}\right)^2(C_{V_i}^2-C_{A_i}^2)\Bigg]\label{iave}\\
I_1^i&=&\frac{32G_F^2\pi p_\nu^3}{15}\Bigg[12C_{V_i}C_{A_i}\left(\frac{p_\nu}{E_i}\right)+(C_{V_i}^2+C_{A_i}^2)
    \left\{ 5 + \left(\frac{p_i}{E_i}\right)^2+\frac{12}{7}\left(\frac{p_\nu}{E_i}\right)^2\right\}\nonumber\\
    &&-3\left(\frac{m_i}{E_i}\right)^2(C_{V_i}^2-C_{A_i}^2)\Bigg]\label{i1}\\
I_2^i&=&\frac{128G_F^2\pi p_\nu^5}{35}\Bigg[20C_{V_i}C_{A_i}\left(\frac{p_\nu}{E_i}\right)+(C_{V_i}^2+C_{A_i}^2)
    \left\{ 7 + \left(\frac{p_i}{E_i}\right)^2+\frac{10}{3}\left(\frac{p_\nu}{E_i}\right)^2\right\}\nonumber\\
    &&-5\left(\frac{m_i}{E_i}\right)^2(C_{V_i}^2-C_{A_i}^2)\Bigg]\label{i2}~.
\end{eqnarray}}
Here $<|M^2|>$ is the squared matrix element summed over final spins and 
averaged over initial spins for a scattering process and $C_V$
and $C_A$ are vector and axial vector coupling constants. 

For non-relativistic nucleons $(m_i/E_i)\simeq1$, $(p_i/E_i)\ll1$ and if also 
$(p_\nu/E_i)\ll 1$, $\lambda_\eta^i$ reduces to \cite{Good}
\begin{equation}
 \lambda_\eta^i=\frac{\frac{11}{35}C_{V_i}^2+\frac{2}{7}C_{A_i}^2}{C_{V_i}^2+2C_{A_i}^2}
\end{equation}
However, we do not assume non-relativistic approximation in this calculation.
The total shear viscosity is given by 
\begin{equation}
\eta_{total} = \eta_n + \eta_p + \eta_e + \eta_{\nu}~,
\end{equation}
where 
\begin{equation}
\eta_{i(=n,p,e)} = \frac {n_i p_{F_i}^2 \tau_i}{5m_i^*}~.
\end{equation}
The relaxation time of i-th species ($\tau_i$) is calculated using Ref. 
\cite{Yak,Bani6}. Effective mass and Fermi momentum of i-th particle
species are denoted by $m_i^*$ and $p_{F_i}$, respectively.
The electron effective mass is taken as its chemical potentials due to 
relativistic effects. 

We are interested in a first order phase transition driven by the nucleation of
droplets of antikaon condensed phase in the neutrino-trapped nuclear matter.
Droplets of antikaon condensed phase are born in the metastable nuclear matter 
due to thermal fluctuations. Droplets of antikaon condensed matter above a 
critical size ($R_c$) will grow and drive the phase transition. According to 
the homogeneous nucleation formalism of Langer and others, the thermal 
nucleation per unit time per unit volume is given by \cite{Lan,Tur}
\begin{equation}
\Gamma = \Gamma_0 exp\left(-\frac {\triangle F (R_c)}{T}\right)~,
\end{equation}  
where $\triangle F$ is the free energy cost to produce a droplet with a 
critical size in the metastable nuclear matter. The free energy shift of the
system as a result of the formation of a droplet is given by \cite{Hei,Min}
\begin{equation}
\triangle F (R) = -\frac{4\pi}{3} (P^K - P^N) R^3 + 4\pi \sigma R^2~,
\end{equation}
where $R$ is the radius of the droplet, $\sigma$ is surface tension of the
interface separating two phases and $P^N$ and $P^K$ are the pressure in 
neutrino-trapped nuclear and antikaon condensed phases, respectively as 
discussed above. We 
obtain the critical radius of the droplet from the maximum of $\triangle F(R)$ 
i.e.  $\delta_R \triangle F = 0$, 
\begin{equation}
R_C = \frac{2 \sigma}{(P^K - P^N)}~.
\end{equation}
This relation also demonstrates  the mechanical equilibrium between two phases. 

We write the prefactor in Eq. (18) as the product of two parts - statistical and
dynamical prefactors \cite{Tur,Las,Raj}
\begin{equation}
\Gamma_0 = \frac{\kappa}{2\pi} \Omega_0~. 
\end{equation}
The available phase space around the saddle point at $R_C$ during the passage 
of the droplet through it is given by the statistical prefactor ($\Omega_0$), 
\begin{equation}
\Omega_0 = \frac{2}{3\sqrt{3}} \left(\frac {\sigma}{T}\right)^{3/2} 
\left(\frac {R_C}{\xi}\right)^4~,
\end{equation}
Here $\xi$ is the kaon correlation length which is considered to be the width of
the interface between nuclear and antikaon condensed matter. 
The dynamical prefactor $\kappa$ is responsible for
the initial exponential growth rate of a critical droplet and given by 
\cite{Las,Raj}
\begin{equation}
\kappa = \frac {2\sigma}{R_C^3 (\triangle w)^2} \left[ \lambda T + 2 
(\frac{4}{3} \eta + \zeta)\right]~.
\end{equation}
Here $\triangle w = w_{K} - w_{N}$ is the enthalpy difference between two 
phases,
$\lambda$ is the thermal conductivity and $\eta$ and $\zeta$ are 
the shear and bulk viscosities of neutrino-trapped nuclear matter. 
We neglect the contribution of thermal conductivity because it is smaller 
compared with that of shear viscosity \cite{Bani7}. We also do not consider 
the contribution of bulk viscosity in the prefactor in this calculation.

We can now calculate the thermal nucleation time ($\tau_{nuc}$) in the interior 
of neutron stars as
\begin{equation}
\tau_{nuc} = \left(V \Gamma \right)^{-1}~,
\end{equation}
where the volume $V = 4{\pi}/3 R_{nuc}^3$. We assume that pressure and
temperature are constant within this volume in the core. 

\section{Results and Discussion}
Nucleon-meson coupling constants in this calculation are taken from Glendenning
and Moszkowski parameter set known as GM1 \cite{Gle91} 
and those are obtained by reproducing the saturation properties of
nuclear matter such as binding energy $E/B=-16.3$ MeV, baryon density 
$n_0=0.153$ fm$^{-3}$, asymmetry energy coefficient $a_{\rm asy}=32.5$ MeV, 
incompressibility $K=300$ MeV and effective nucleon mass $m^*_N/m_N = 0.70$. 

Next kaon-meson vector coupling constants are estimated from 
the quark model and isospin counting rule \cite{Gle99,Bani6}. The scalar 
kaon-meson coupling constant is obtained from the real part of $K^-$ optical 
potential depth at normal nuclear matter density.
The strength of antikaon optical potential depth is obtained from 
heavy ion collision experiments and $K^-$ atomic data 
\cite{Fri94,Fri99,Koc,Waa,Li,Pal2}. It has a wide range of values. 
On the one hand, the analysis of $K^-$ 
atomic data predicted the real part of the antikaon optical potential to be as 
large as $U_{\bar K} (n_0) = -180 \pm 20$ MeV at normal nuclear matter density
\cite{Fri94,Fri99}. On the other hand, theoretical models including chirally 
motivated coupled channel models as well as a double pole structure of $\Lambda 
(1405)$ yielded a less attractive antikaon optical potential depth 
\cite{Ram,Koch,Mag,Hyo}. Here we consider an antikaon optical potential depth 
of $U_{\bar K} (n_0) = -120$ MeV at normal nuclear matter density and the 
corresponding kaon-scalar meson coupling constant is $g_{\sigma K} = 1.6337$. 
The value of antikaon optical potential adopted in this calculation resulted in
a maximum neutron star mass of 2.08 $M_{\odot}$ in earlier calculations
using the Maxwell construction \cite{Pal}. This is consistent with the recently 
observed 2$M_{\odot}$ neutron star \cite{Demo}.  

Using the above mentioned nucleon-meson and kaon-meson coupling constants, we
calculate the EoS of neutrino-trapped nuclear and antikaon condensed phases at 
zero temperature in a self-consistent manner. A temperature of a few tens of 
MeV might modify the EoS at very high densities in (proto)neutron stars 
compared with the zero temperature EoS as it was noted in Ref.\cite{Pons}. However,
the effect of finite temperature on the threshold of antikaon condensation is negligible
\cite{Pons}. In this calculation, the EoS enters in Eq. (19) as the difference 
between  pressures
in two phases and in Eq. (23) as the enthalpy difference between two phases.  
Here we exploit the zero temperature EoS for the
the calculation of shear viscosity and thermal nucleation time. The thermal nucleation of exotic
phases was earlier investigated using zero temperature EoS in Ref.\cite{Bani7,Min}. 
 
First we calculate shear viscosities of neutrons, protons and electrons in
neutrino-trapped nuclear matter using Eq. (5) in the same fashion as it was 
done in Ref.\cite{Bani7}. We take lepton fraction $Y_L = 0.4$ in this 
calculation. Shear viscosities of different species are shown as a 
function of normalised baryon density at a temperature T=10 MeV in Fig. 1.
Here the electron viscosity is higher than the neutron and proton shear 
viscosities. The total shear viscosity excluding the
viscosity due to neutrinos in neutrino-trapped nuclear matter is shown as a 
function of normalised baryon density for temperatures $T =$ 1, 10, 30 and 100 
MeV in Fig. 2. The shear viscosity is found to increase with baryon density. 
Furthermore, the shear viscosity decreases with increasing temperature. 
The temperature dependence of shear viscosities is a complex one and quite 
different from the characteristic $1/T^2$ behaviour of a Fermi liquid 
\cite{Yak}. It is observed that the shear viscosities of neutrons, protons and
electrons in neutrino-trapped nuclear matter are of the same orders of 
magnitude as those of the neutrino-free case \cite{Bani6}. 

Next we calculate the shear viscosity due to neutrinos. As a prelude to it, we 
compare effective
relaxation times corresponding to different species in neutrino-trapped nuclear
matter in Fig. 3. Relaxation time is plotted with normalised baryon density at
a temperature T=10 MeV in Fig. 3. Relaxation times of different particle 
species due to scattering under strong and electromagnetic interactions are much
much smaller than that of neutrinos undergoing scattering with other species 
through weak interactions. Consequently, particles excluding neutrinos come into thermal 
equilibrium quickly on the time scale of weak interactions. We 
calculate the shear viscosity due to neutrinos only treating others as 
background and it is shown as a 
function of normalised baryon density for temperatures $T =$ 1, 10, 30 and 100
in Fig. 4. Like Fig. 2, the neutrino shear viscosity decreases with increasing 
temperature. However, the neutrino shear viscosity is several orders of 
magnitude larger than shear viscosities of neutrons, protons and electrons 
shown in Fig. 2. It is the neutrino shear viscosity which dominates the total
viscosity of Eq. (16) in neutrino-trapped matter. We perform the rest of our 
calculation using the neutrino viscosity in the following paragraphs.

We calculate the prefactor ($\Gamma_0$) according to Eqs. (21)-(23). The 
dynamical prefactor not only depends on the shear viscosity but also on the 
thermal
conductivity and bulk viscosity. However, it was already noted that the thermal 
conductivity and bulk viscosity in neutrino-trapped nuclear matter were 
negligible compared with the shear viscosity \cite{Good}. We only consider the
effect of shear viscosity on the prefactor. Besides transport coefficients, the
prefactor in particular, the statistical prefactor is sensitive to the 
correlation length of kaons and surface tension. The correlation length is 
the thickness of the interface between nuclear and kaon phases 
\cite{Bom,Las} having a value $\sim$ 5 fm \cite{San}. 
The radius of a critical droplet is to be greater than the correlation length 
($\xi$) for
kaons \cite{San,Las}. We perform our calculation with antikaon droplets 
with radii greater than 5 fm. The other important parameter in the prefactor is
the surface tension. The surface tension between nuclear and kaon phases was 
already estimated by Christiansen and collaborators \cite{Chri} and found to be
sensitive to the EoS. 
We perform this calculation for a set of values of surface 
tension $\sigma =$ 15, 20, 25 and 30 MeV fm$^{-2}$. The prefactor ($\Gamma_0$) 
is shown as a function of temperature in Fig. 5. It is shown for a baryon 
density $n_b = 4.235n_0$ which is just above the critical density 3.9$n_0$ for
antikaon condensation at zero temperature \cite{Bani1}, and 
surface tension $\sigma = 15$ MeV fm$^{-2}$. 
The prefactor was also approximated by $T^4$ according to the dimensional 
analysis in many calculations \cite{Las,Min}. The upper curve in Fig. 5 shows 
the prefactor of Eq. (21) including only the contribution of neutrino shear 
viscosity whereas the prefactor approximated by
$T^4$ corresponds to the lower curve. It is evident from Figure 5 that the 
approximated prefactor is very small compared with our result.

Now we discuss the nucleation time of a critical droplet of antikaon condensed 
phase in neutrino-trapped nuclear matter and the effect of neutrino shear 
viscosity on it. The thermal nucleation rate of the critical droplet is
calculated within a volume with $R_{nuc}=100$ m in the core of a neutron star 
where the density, pressure and temperature are constant. The thermal 
nucleation time is plotted 
with temperature for a baryon density $n_b = 4.235n_0$ in Fig. 6. 
Furthermore, this 
calculation is done with the kaon correlation length $\xi =$ 5 fm and surface 
tension $\sigma =$ 15, 20, 25 and 30 MeV fm$^{-2}$. The size of the critical 
droplet increases with increasing surface tension. Radii of the critical 
droplets are 7.1, 9.4, 11.7 and 14.1 fm corresponding to $\sigma=$ 15, 20, 25 
and 30 MeV fm$^{-2}$, respectively, at a baryon density $4.235n_0$. The 
nucleation time of the critical droplet diminishes as temperature increases for
all cases studied here. However, the temperature corresponding to a particular 
nucleation time for example $10^{-3}$ s, increases as the surface tension increases. 
There is a possibility that the condensate might melt down if the temperature 
is higher than the critical temperature. So far, there is no calculation of critical temperature
of antikaon condensation in neutrino-trapped matter. However, the critical temperature of antikaon
condensation was investigated in neutrino-free matter in Ref.\cite{Bani5}. 
We compare thermal nucleation times corresponding to different values of the surface tension with 
the early post bounce time scale $t_{d} \sim$ 100 ms in the core collapse supernova 
\cite{Sag} 
when the central density might reach the threshold density of antikaon condensation. The 
time scale $t_d$ is much less than the neutrino diffusion time $\sim$ 1 s as obtained by 
Ref.\cite{Good}. Thermal nucleation of the antikaon condensed phase may be possible when the 
thermal nucleation time is less than $t_{d}$. For   
$\sigma =$ 15 MeV fm$^{-2}$, the thermal nucleation time of $10^{-3}$ s occurs at a 
temperature 16 MeV. It is evident from 
Fig. 6 that the thermal nucleation time is strongly dependent on the surface tension. 
Further thermal nucleation of an antikaon droplet is possible so long as   
the condensate might survive the melt down at high temperatures\cite{Bani5}. 
Our results of thermal nucleation time are compared with the calculation taking
into account the prefactor approximated by $T^4$ in Fig. 7 for surface tension 
$\sigma =$ 15 MeV fm$^{-2}$ and at a density $n_b = 4.235n_0$. The upper curve
denotes the calculation with $T^4$ approximation whereas the lower curve
corresponds to the influence of neutrino shear viscosity on the thermal 
nucleation time. The results of the $T^4$ approximation overestimate our 
results hugely. For a nucleation time of $10^{-3}$ s at a temperature T=16 Mev, the
corresponding time in the $T^4$ approximation is larger by several orders of
magnitude.

\section{Summary and Conclusions}
We have studied shear viscosities of different particle species in 
neutrino-trapped $\beta$-equilibrated and charge neutral nuclear matter. We 
have derived equations of state of nuclear
and antikaon condensed phases in the relativistic mean field model for the 
calculation of shear viscosity. It is noted that neutrons, protons and 
electrons come into thermal equilibrium in the weak interaction time scale.
The
shear viscosity due to neutrinos is calculated treating other particles as
background and found to dominate the total shear viscosity.
 
Next we have investigated the first-order phase transition from 
neutrino-trapped nuclear matter to antikaon condensed matter through the 
thermal nucleation of a critical droplet of antikaon condensed matter using
the same relativistic EoS as discussed above. 
Our emphasis in this calculation is the role of the shear viscosity due to 
neutrinos in the prefactor and its consequences on the thermal 
nucleation rate. We have observed that the thermal nucleation 
of a critical antikaon droplet might be possible well before the neutrino diffusion 
takes place. Furthermore, a comparison of our 
results with that of the calculation of thermal nucleation time in the $T^4$ 
approximation shows that the latter overestimates our results of thermal 
nucleation time computed with the prefactor including the neutrino shear 
viscosity. Though we have performed
this calculation with antikaon optical potential depth of 
$U_{\bar K} (n_0) = -120$ MeV, we expect qualitatively same results for other 
values of the antikaon optical potential depth \cite{Bani8}.

\newpage
\vspace{-2cm}

{\centerline{
\epsfxsize=12cm
\epsfysize=14cm
\epsffile{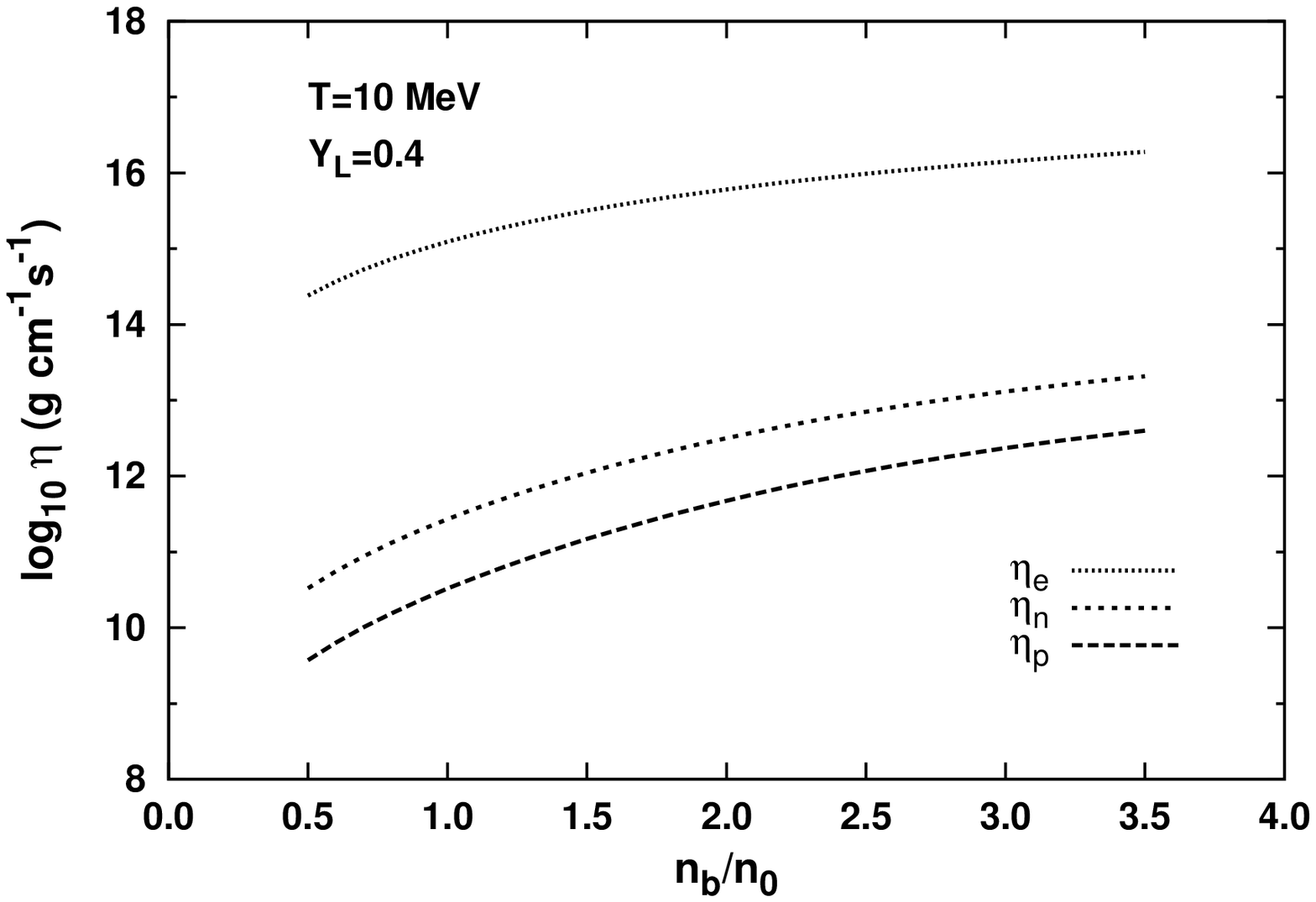}
}}

\vspace{1.0cm}

\noindent{\small{Fig. 1. 
Shear viscosities corresponding to different particle species
in neutrino-trapped nuclear matter are shown as a function of normalised baryon
density at a temperature $T = 10$ MeV and $Y_L=0.4$.}}

\newpage 
\vspace{-2cm}

{\centerline{
\epsfxsize=12cm
\epsfysize=14cm
\epsffile{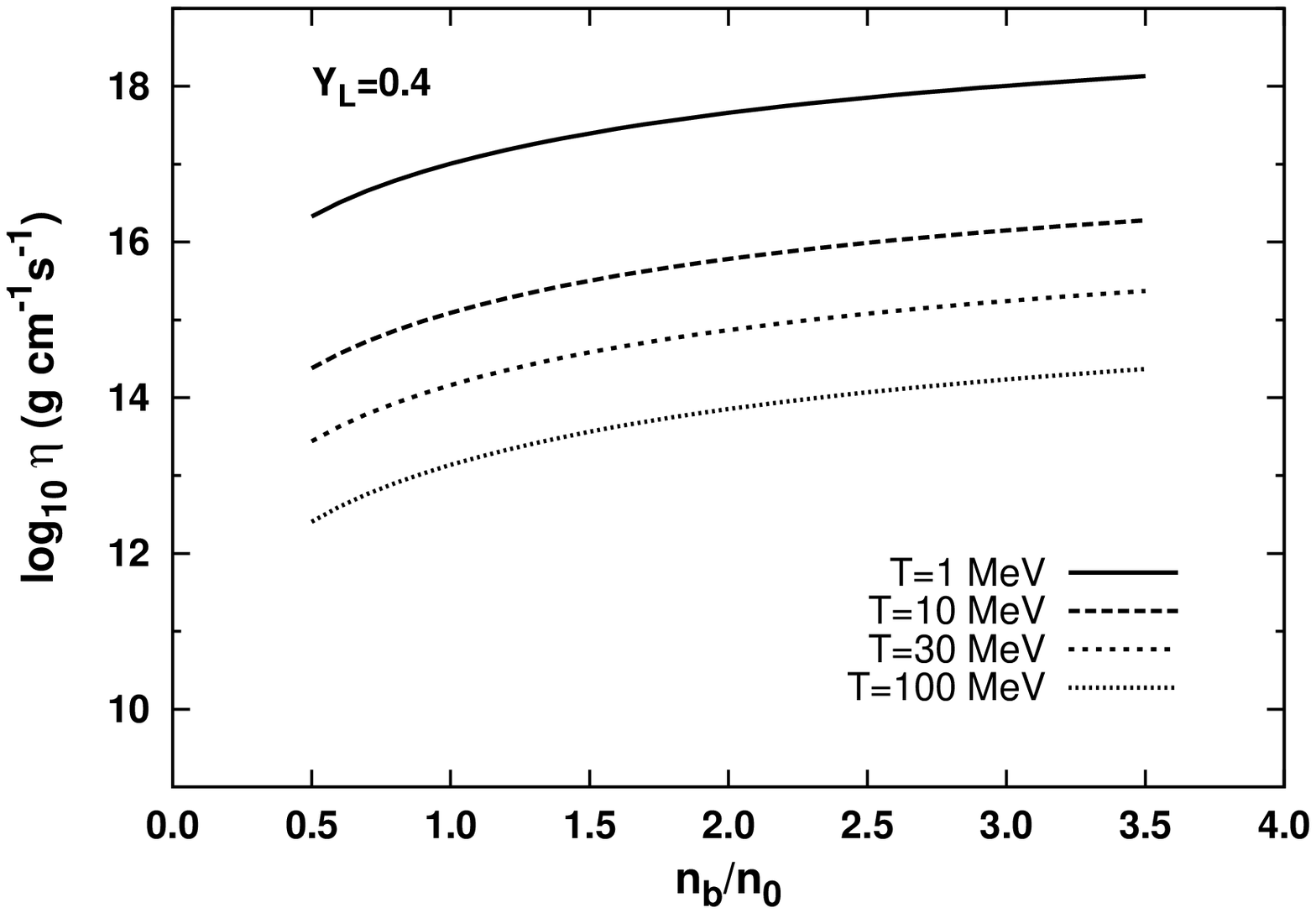}
}}

\vspace{4.0cm}

\noindent{\small{
FIG. 2. Total shear viscosity in neutrino-trapped nuclear matter except the
contribution of neutrinos is plotted with normalised baryon density for 
different temperatures.}}

\newpage
\vspace{-2cm}

{\centerline{
\epsfxsize=12cm
\epsfysize=14cm
\epsffile{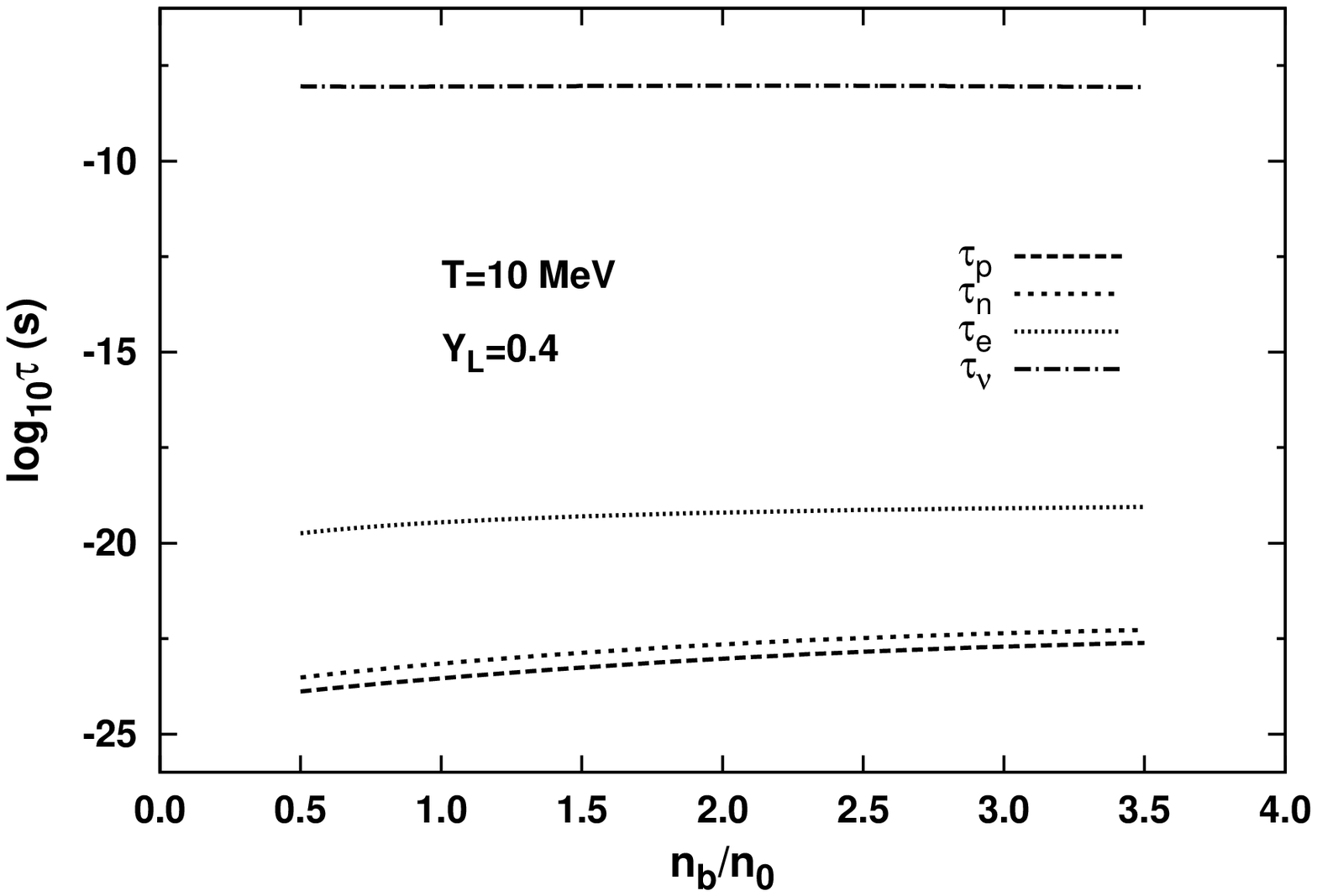}
}}

\vspace{1.0cm}

\noindent{\small{Fig. 3. Relaxation times corresponding to different species
in neutrino-trapped nuclear matter are shown as a function of normalised baryon 
density at a temperature $T = 10$ MeV and $Y_L = 0.4$.}}

\newpage 
\vspace{-2.0cm}

{\centerline{
\epsfxsize=12cm
\epsfysize=14cm
\epsffile{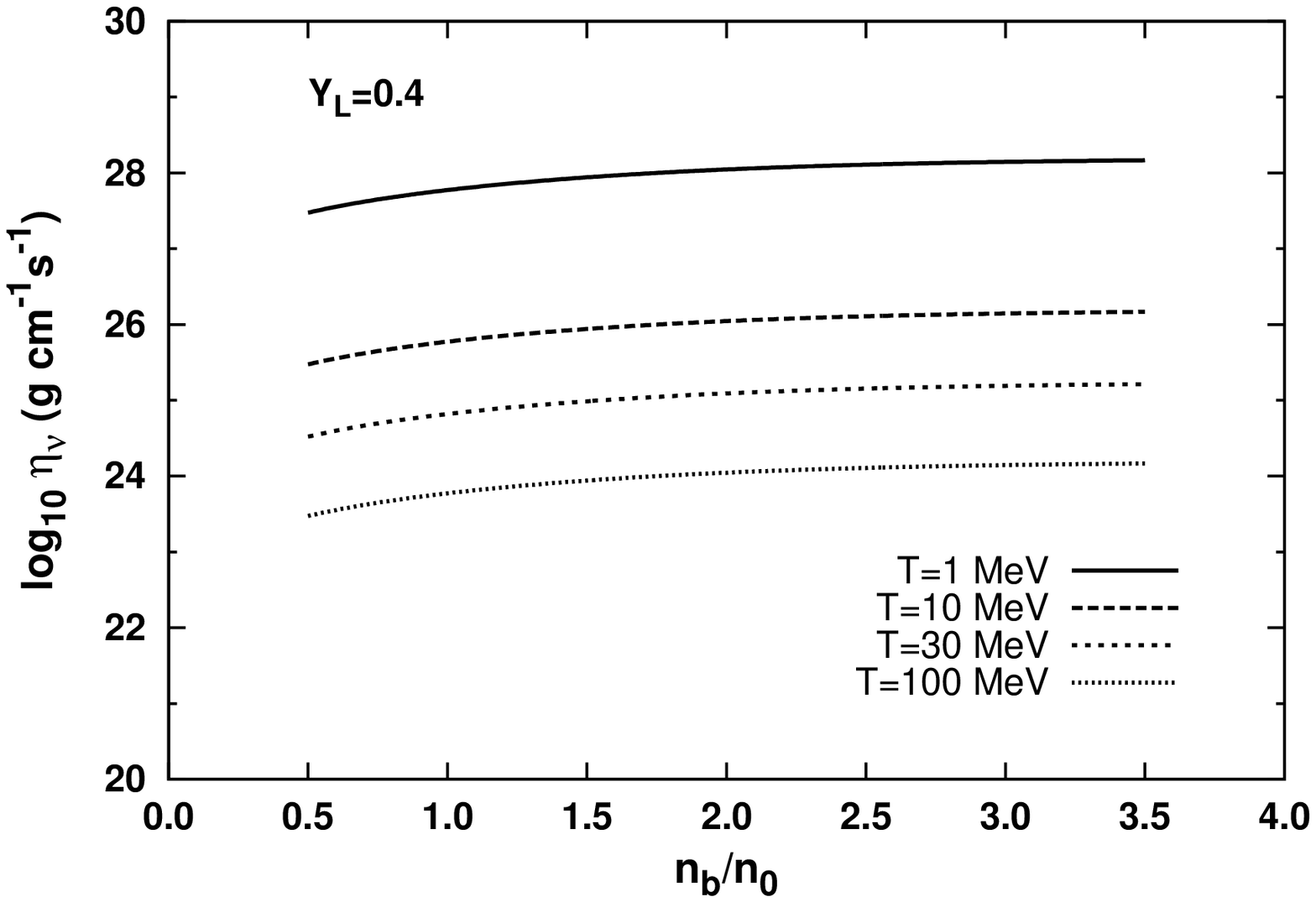}
}}

\vspace{1.0cm}

\noindent{\small{Fig. 4. Neutrino shear viscosity is shown as a function of 
normalised baryon density at different temperatures.}}

\newpage 
\vspace{-2cm}

{\centerline{
\epsfxsize=12cm
\epsfysize=14cm
\epsffile{pre_120bw.eps}
}}

\vspace{4.0cm}

\noindent{\small{
FIG. 5. Prefactor including the contribution of shear
viscosity is plotted as a function of temperature at a fixed baryon density and
surface tension and compared with that of $T^4$ approximation.}}

\newpage 
\vspace{-2cm}

{\centerline{
\epsfxsize=12cm
\epsfysize=14cm
\epsffile{tau_120bw.eps}
}}

\vspace{4.0cm}

\noindent{\small{
FIG. 6. Thermal nucleation time is displayed with temperature for different
values of surface tension.}}

\newpage 
\vspace{-2cm}

{\centerline{
\epsfxsize=12cm
\epsfysize=14cm
\epsffile{tau_thbw.eps}
}}

\vspace{4.0cm}

\noindent{\small{
FIG. 7. Same as Fig.6 but our results for a fixed surface tension are compared 
with the calculation of $T^4$ approximation.}}

\end{document}